\documentclass[12pt]{article}
\usepackage{sbc-template}
\usepackage{graphicx,url}
\usepackage[utf8]{inputenc}
\usepackage[brazil]{babel}
\usepackage[portuguese,algoruled,longend,commentsnumbered]{algorithm2e/algorithm2e}
\usepackage{algpseudocode}
\usepackage{enumerate}
\usepackage{amsmath}
\usepackage{amsthm}
\usepackage{amssymb}
\usepackage{array}
\usepackage{longtable}
\usepackage{listings}
\usepackage{multirow}
\usepackage{pslatex}
\usepackage{geometry} 
\usepackage{setspace}
\usepackage{epstopdf}
\usepackage{colortbl}
\definecolor{green(munsell)}{rgb}{0.0, 0.66, 0.47}
\definecolor{lightseagreen}{rgb}{0.13, 0.7, 0.67}
\definecolor{mint}{rgb}{0.24, 0.71, 0.54}
\definecolor{persiangreen}{rgb}{0.0, 0.65, 0.58}
\definecolor{DarkTurquoise}{RGB}{0,206,209}
\definecolor{LightSeaGreen}{RGB}{32,178,170}
\definecolor{darkred}{RGB}{188, 0, 28}
\definecolor{LightBlue}{RGB}{67, 184, 244}
\definecolor{MidnightBlue}{RGB}{25, 25, 112}
\definecolor{Snow2}{RGB}{238, 233, 233}
\definecolor{Snow3}{RGB}{205, 201, 201}
\definecolor{PinkLight}{RGB}{255, 225, 225}
\definecolor{BlueLight1}{RGB}{235, 235, 255}
\definecolor{BlueLight2}{RGB}{225, 235, 255}

\DeclareGraphicsExtensions{.pdf,.eps,.png,.jpg,.mps}
\usepackage{float} 
\usepackage{comment}

\title{Estudo comparativo de meta-heurísticas para problemas de colorações de grafos\footnote{Projeto de Produtividade Acadêmica da Universidade do Estado do Amazonas (UEA), processo n\textordmasculine~347418, portaria n\textordmasculine~1112/2014.}}

\author{Flávio José Mendes Coelho\inst{1}}

\address{Curso de Engenharia de Computação --  Escola Superior de Tecnologia  \\ Universidade do Estado do Amazonas (UEA)\\
  Manaus -- AM -- Brazil
\email{\{fcoelho\}@uea.edu.br}
}

\begin{document} 

\maketitle

\begin{abstract}
A classic graph coloring problem is to assign colors to vertices of any graph so that distinct colors are assigned to adjacent vertices. Optimal graph coloring colors a graph with a minimum number of colors, which is its chromatic number. Finding out the chromatic number is a combinatorial optimization problem proven to be computationally intractable, which implies that no algorithm that computes large instances of the problem in a reasonable time is known. For this reason, approximate methods and metaheuristics form a set of techniques that do not guarantee optimality, but obtain good solutions in a reasonable time. This paper reports a comparative study of the Hill-Climbing, Simulated Annealing, Tabu Search, and Iterated Local Search metaheuristics for the classic graph coloring problem considering its time efficiency for processing the DSJC125 and DSJC250 instances of the DIMACS benchmark.\end{abstract}
     
\begin{resumo} 
Um problema clássico de coloração de grafos consiste em atribuir-se cores a vértices de um grafo qualquer de forma que sejam atribuídas cores distintas a vértices adjacentes. Uma coloração de grafo ótima colore um grafo com um número  mínimo de cores, sendo este o seu número cromático. Descobrir o número cromático é um problema de otimização combinatória provado ser computacionalmente intratável, o que implica não ser conhecido qualquer algoritmo que compute instâncias grandes do problema em tempo razoável. Por esta razão, métodos aproximados, heurísticas e meta-heurísticas formam um conjunto de técnicas que não garantem alcançar otimalidade, mas obtém boas soluções em tempo razoável. Este artigo relata um estudo comparativo entre as meta-heurísticas \textit{Hill-Climbing}, \textit{Simulated Annealing}, \textit{Tabu Search} e \textit{Iterated Local Search} para o problema clássico de coloração de grafos considerando sua eficiência de tempo para o processamento das instâncias DSJC125 e DSJC250 do benchmark DIMACS.
\end{resumo}

\section{Introdução}

Um grafo reúne um conjunto de elementos chamados de \emph{vértices} e outro conjunto cujos elementos denominam-se \emph{arestas} que formam pares não ordenados de vértices. A partir desta estrutura combinatória simples é possível modelar uma miríade de aplicações teóricas e práticas em um amplo espectro de áreas do conhecimento tais como biologia, linguística, engenharias e computação. Um dos problemas clássicos em teoria dos grafos consiste em se colorir os vértices de um grafo com um número mínimo de cores, considerando a restrição de que vértices adjacentes (formando aresta) não assumam uma mesma cor. Problemas de coloração de grafos modelam relevantes problemas práticos tais como programação de horários em universidades e escolas \cite{Diestel:2005}, alocação de canais em redes de telecomunicação \cite{Hale:1980}, projeto de armazenamento seguro de produtos químicos incompatíveis, programação de períodos de exames em universidades \cite{Bondy:2008} e coloração de mapas geopolíticos. Estes problemas são computacionalmente intratáveis o que torna altamente custoso a obtenção de soluções ótimas para instâncias reais (grandes), ainda que se disponha de uma plataforma computacional robusta para a sua resolução. Desta forma, organizam-se dois conjuntos de técnicas para esta classe de problemas: as técnicas exatas que garantem soluções ótimas, mas demandam um alto custo de recursos como tempo de processamento e memória, devido à intratabilidade computacional; e, as técnicas heurísticas que trocam a garantia de otimalidade pela obtenção de soluções de qualidade aceitável, consumindo tempo ou espaço de forma razoável. Dentre as técnicas heurísticas, encontram-se as meta-heurísticas que se inspiram em modelos de sistemas biológicos, físicos, naturais, entre outros, combinados com técnicas estocásticas. Isto confere algum grau de não determinismo aos procedimentos empregados por meta-heurísticas, o que requer um trabalho empírico de ajustes de parâmetros destes procedimentos para se alcançar soluções aceitáveis.

Este trabalho apresenta um estudo comparativo da eficiência de tempo entre as meta-heurísticas \textit{Hill-Climbing}, \textit{Simulated Annealing}, \textit{Tabu Search} e \textit{Iterated Local Search} para o problema clássico de coloração de grafos.

\section{Fundamentação teórica}\label{sec:fundamentacao}

Esta seção aborda conceitos relacionados à coloração de grafos, complexidade computacional, otimização combinatória e apresenta as meta-heurísticas utilizadas neste trabalho.

\subsection{Coloração de grafos}

Um \emph{grafo simples} (ou, apenas grafo) é um par ordenado $G = (V, E)$, onde $V$ é um conjunto não-vazio e $E$ é um conjunto de pares não-ordenados de elementos de $V$. Os conjuntos $V$ e $E$ são finitos, disjuntos, e seus elementos são \emph{vértices} e \emph{arestas}, respectivamente. Considera-se $n = |V|$ a \emph{ordem} de um grafo e $m = |E|$ seu \emph{tamanho}. 
Uma aresta $e = \{u, v\} \in E$ une dois vértices $u, v \in V$ que são seus \emph{extremos} ou extremidades. Os vértices $u$ e $v$ de uma aresta são ditos \emph{adjacentes} entre si \cite{Bondy:2008}. 

Uma \emph{coloração de vértices} de um grafo $G = (V, E)$ é uma função $f:V \rightarrow C$ que associa cada vértice de $G$ a uma cor de um conjunto de cores $C \subset \mathbb{Z}^{+}$. De outra forma, uma coloração é uma atribuição de cores aos vértices de um grafo $G$. Uma coloração é \emph{própria} se $f(u) \neq f(v)$ para qualquer aresta $\{u, v\} \in E$, isto é, vértices adjacentes devem ser assinalados com cores distintas. Uma \emph{$k$-coloração} é uma coloração $f$ tal que $f(v) \leq k$ para cada $v \in V$, e um grafo é \emph{$k$-colorível} se tem uma $k$-coloração. O inteiro positivo mínimo $k$ para o qual um grafo $G$ é $k$-colorível é seu \emph{número cromático} $\chi(G)$. Um grafo com $\chi(G) = k$ é denominado \emph{$k$-cromático} \cite{Chartrand:2008:CGT:1457583}. Denomina-se \textsc{número-cromático} o problema de se obter o número cromático de um grafo $G$ qualquer. O conceito de coloração própria associado ao problema do~\textsc{número-cromático} constitui a matriz de numerosas variações de colorações e problemas com as mais diversas aplicações. Coloração de arestas, coloração total \cite{Behzad:1965, Vizing:1964}, \textit{sum coloring} \cite{Kubicka:1989}, lista coloração \cite{Vizing:1976, Erdos:1979}, T-coloração \cite{Hale:1980}, $\mu$-coloração \cite{Bonomo:2011} e  \textit{fractional coloring} \cite{Scheinerman:1997}, são alguns exemplos de tipos de coloração pesquisadas em teoria dos grafos

\subsection{Complexidade computacional}

Para apresentar a dificuldade computacional envolvida na resolução do \textsc{número-cromático} é necessário a introdução de uma teoria denominada de \emph{complexidade computacional} que investiga: (a) a análise da eficiência de algoritmos, em geral, eficiência de tempo ou de espaço; 
(b) as técnicas de projeto de algoritmos mais adequadas ao tipo de problema a ser tratado; (c) e, a classificação de problemas quanto a sua dificuldade computacional \cite{Cormen:2009:IAT:1614191, Sedgewick:2011:ALG:2011916, Levitin:2012:IDAA}. 

Considere a definição de um problema computacional $\Pi$ como consistindo de um conjunto $D_{\Pi}$ de elementos finitos denominados \emph{instâncias} e, para qualquer instância $I \in D_{\Pi}$, um  conjunto $S_{\Pi}(I)$ de elementos finitos chamados de \emph{soluções} de $I$ \cite{Cormen:2009:IAT:1614191}. Um \emph{problema de decisão} é uma questão do tipo SIM/NÃO sobre a existência de um objeto que satisfaça a um conjunto de parâmetros do problema. De outro modo, dada uma instância $I \in D_{\Pi}$ de um problema de decisão $\Pi$, $S_{\Pi}(I) = \{\text{``SIM''}\}$ se existir tal objeto, ou $S_{\Pi}(I) = \{\text{``NÃO''}\}$, em caso contrário. Um \emph{problema de localização} (ou de busca) solicita a apresentação de um objeto que satisfaça a um conjunto de parâmetros do problema, caso tal objeto exista. Dessa forma, se $I \in D_{\Pi}$ é uma instância de um problema de localização $\Pi$ e $Q$ é o objeto solicitado pela instância do problema, então $S_{\Pi}(I) = \{Q\}$, se $Q$ existir, ou $S_{\Pi}(I) = \emptyset$, em caso contrário. Embora possa existir um conjunto de objetos que satisfaçam os parâmetros do problema, qualquer um dos objetos é suficiente como solução. Enfim, um \emph{problema de otimização} solicita a apresentação de um objeto que satisfaça a um conjunto de parâmetros do problema, e que atenda a algum critério de maximização ou de minimização \cite{Cormen:2009:IAT:1614191}. É fácil perceber que um problema de otimização embute um problema de localização que, por sua vez, embute um problema de decisão. Por esse fato, um mesmo problema pode ser tratado do ponto de vista de decisão, de localização ou de sua versão de otimização.

Diz-se que um algoritmo \emph{resolve} um problema de decisão $\Pi$ se consegue mapear uma instância qualquer de $\Pi$ em uma solução-SIM ou em uma solução-NÃO, e então para. Um algoritmo \emph{resolve um problema em tempo} $O(f(n))$, se para cada instância $I \in D_{\Pi}$ de tamanho $n$ do problema $\Pi$, produz uma solução em um número proporcional a $f(n)$ passos de execução. O algoritmo é \emph{polinomial} se $f(n) = n^k$, para algum inteiro não negativo $k$, isto é, seu tempo pode ser expresso como um polinômio $p(n)$. Em contraste, um \emph{algoritmo exponencial} tem $f(n) = k^n$. Um problema de decisão $\Pi$ está na \emph{classe de complexidade P} ($\Pi \in \text{P}$) ou é dito ser \emph{tratável}, se existe um algoritmo polinomial para $\Pi$ \cite{Cormen:2009:IAT:1614191, Levitin:2012:IDAA}.

Dada uma instância $I$ de um problema de decisão $\Pi$, um \emph{certificado} para uma solução-SIM de $I$ é uma solução correspondente à versão de localização de $\Pi$. Diz-se que uma solução-SIM da instância $I$ é \emph{verificada} (provada ser verdadeira), se existe um algoritmo que toma como entrada o par ($I$, $C$), onde $C$ é um certificado para $S_{\Pi}(I) = \{\text{``SIM''}\}$, e tem como saída um SIM, indicando que $C$, de fato, é uma solução não vazia da versão de localização de $\Pi$. Desse modo, um problema de decisão $\Pi$ está na \emph{classe de complexidade NP} ($\Pi \in \text{NP}$) ou é dito ser \emph{intratável}, se todas as suas soluções podem ser verificadas em tempo polinomial. É importante notar que a verificação em tempo polinomial não implica na resolução do problema em tempo polinomial\footnote{Porém, obter um certificado a partir de uma instância do problema requer um algoritmo exponencial.}, embora o contrário seja verdadeiro (resolução polinomial implica em verificação polinomial). Um problema $\Pi$ está na classe \emph{NP-completo} se está em NP e se todos os problemas em NP podem ser transformados em $\Pi$ em tempo polinomial \cite{Cormen:2009:IAT:1614191}.

No problema de decisão chamado \textsc{K-Coloribilidade}, pergunta-se: dado um grafo qualquer $G = (V, E)$, existe um inteiro positivo $k \leq |V|$ para o qual $G$ é $k$-colorível? O \textsc{Número-Cromático} é a versão de otimização da \textsc{K-Coloribilidade}, pois solicita o $k$ mínimo para o qual $G$ é $k$-colorível. Como a \textsc{K-Coloribilidade} pertence à classe NP-completo \cite{GareyJohnson:1979}, por consequência, o \textsc{Número-Cromático} é tão difícil quanto a \textsc{K-Coloribilidade}. Desta forma, para um grafo qualquer o \textsc{Número-Cromático} só pode ser computado deterministicamente por meio de um algoritmo exponencial (a menos que se prove que P = NP \cite{Clay:2019}). Na prática, isto significa que instâncias maiores do problema tem um consumo de tempo proibitivo, pois para o crescimento linear do tamanho da instância do problema, o tempo do algoritmo cresce exponencialmente. A intratabilidade de problemas como o \textsc{Número-Cromático} conduziu à investigação de técnicas que não garantem obter uma solução ótima para o problema, mas obtém uma solução aproximadamente ótima em tempo razoável. Estas técnicas aproximativas se reúnem sob o nome de heurísticas e meta-heurísticas.

\subsection{Otimização combinatória e meta-heurísticas}

Uma \emph{instância de um problema de otimização combinatória} é um par $(F, f)$, onde $F$ é um conjunto qualquer (domínio de pontos viáveis) e $f: F \longrightarrow \mathbb{R}$ é uma \emph{função objetivo} (ou de custo). A questão é achar $x^{*} \in F$ tal que $f(x^{*}) \leq f(x)$, para todo $x \in  F$. Diz-se que $x^{*}$ é uma \emph{solução ótima global} para a instância. Em geral, $F$ é finito e muito grande, ou infinito e contável, e seus elementos são estruturas discretas tais como inteiros, conjuntos, permutações e grafos. Um \emph{problema de otimização combinatória} (POC) é um conjunto $I$ de instâncias de problemas de otimização \cite{Papadimitriou:1982:COA:31027}. De outra forma, em um problema de otimização combinatória busca-se um elemento de $F$ cujo custo (mínimo ou máximo) satisfaça a função objetivo definida sobre o problema. Há POCs cujas soluções são obtidas por algoritmos polinomiais. Porém, há POCs cujas versões de decisão são NP-completo. O \textsc{Número-Cromático} tratado neste trabalho é um destes problemas. 

Pode-se classificar as técnicas para resolução de POCS em: \emph{métodos exatos} e \emph{métodos heurísticos}\footnote{Há, também, os algoritmos aproximados \cite{Feofiloff:2001}, programação semidefinida \cite{Matousek:2013}, entre outras técnicas.}. Os métodos exatos (programação dinâmica, programação inteira, \textit{branch and bound}, relaxação lagrangeana, etc) garantem obter soluções ótimas. Entretanto, a intratabilidade de muitos POCs torna inviável o uso destes métodos, a menos que sejam aplicados a instâncias pequenas do problema em questão. Por outro lado, os métodos heurísticos compreendem um conjunto de técnicas heurísticas que resolvem POCs obtendo uma ou mais soluções de custo satisfatório e em tempo razoável, porém, sem garantir otimalidade \cite{Goldbarg:2017:otimizacao}. Dentre os métodos heurísticos, as \emph{meta-heurísticas} (ou, \emph{métodos estocásticos}) são estratégias generalizadas e estocásticas para POCs que partem de uma solução inicial aleatória e, por meio de um conjunto de operações que geram novas soluções candidatas, obtém uma solução final de boa qualidade, não necessariamente ótima \cite{Luke2013Metaheuristics}. São exemplos de meta-heurísticas: \textit{Hill-climbing} (escalada), \textit{Simulated Annealing} \cite{Kirkpatrick:1983} (recozimento simulado), \textit{Tabu search} \cite{Glover:1986:FPI:15310.15311} (busca tabu), \textit{Iterated Local Search} \cite{Lourenco:2003} (busca local iterada) e \textit{Genetic Algorithms} \cite{Holland:1975, Mitchell:1996:IGA:230231} (algoritmos genéticos), \textit{Ant Colony Optimization} \cite{Dorigo:2006} (otimização por colônia de formigas), \textit{Particle Swarm Optimization} \cite{Eberhart:1995} (otimização por enxame de partículas), entre outras \cite{BOUSSAID201382, Manualmetaheuristicas:104946}. Neste trabalho, as quatro primeiras meta-heurísticas citadas foram empregadas para resolver o \textsc{Número-Cromático}. 

Luke \cite{Luke2013Metaheuristics} classifica meta-heurísticas em \emph{métodos de estado único} (\textit{single state methods}) e \emph{métodos populacionais} (\textit{population-based methods}). Nos métodos de estado único os algoritmos tratam uma única solução candidata a cada passo de otimização. Este é o caso das meta-heurísticas \textit{Hill-climbing}, \textit{Simulated Annealing}, \textit{Tabu search}, \textit{Iterated Local Search}. Nos métodos populacionais, diversas soluções candidatas são manipuladas e transformadas a cada etapa de otimização para alcançar a solução final. Este é o caso de \textit{Genetic Algorithms}. Esta classificação foi utilizada para guiar a modelagem e a codificação das meta-heurísticas.


\section{Método e procedimentos}

Este trabalho se desenvolveu de acordo com as seguintes etapas e procedimentos:

\begin{enumerate}[(1)]
\item Estudo dos fundamentos de teoria de grafos, de problemas de coloração de grafos e assuntos relacionados a estes problemas.
\item Estudo das meta-heurísticas \textit{Hill-climbing} (HC), \textit{Simulated Annealing} (SA), \textit{Tabu search} (TS) e \textit{Iterated Local Search} (ILS).
\item Projeto de experimentos por meio da coleta de instâncias de referência para os problemas de coloração. Foram utilizadas as instâncias de coloração DSJC125 e DSJC250 do II Desafio DIMACS (\textit{Center for Discrete Mathematics and Theoretical Computer Science - Rutgers University}) de Implementação de Cliques, Coloração e Satisfatibilidade de 1993 \cite{Johnson:1996}. Este conjunto de instâncias é referência na literatura de otimização combinatória e serviu tanto para testes de funcionalidade quanto para os experimentos.
\item Codificação das meta-heurísticas do passo (2) em uma linguagem de programação. 
\item Execução de experimentos com as meta-heurísticas codificadas sobre as instâncias de referência.
\item Comparação das meta-heurísticas de acordo com os resultados obtidos nos experimentos.
\end{enumerate}

Os estudos citados nas etapas (1) e (2) foram realizados e apresentados na seção \ref{sec:fundamentacao} deste artigo. O passo (3) foi executado obtendo-se as instâncias na Internet \cite{DIMACS1:1992, DIMACS2:1992, DIMACS3:1992}. Os passos seguintes e seus procedimentos são discutidos nas seções seguintes.



\subsection{Codificação das meta-heurísticas}

\begin{figure}
\centering
\includegraphics[width=1.0\textwidth]{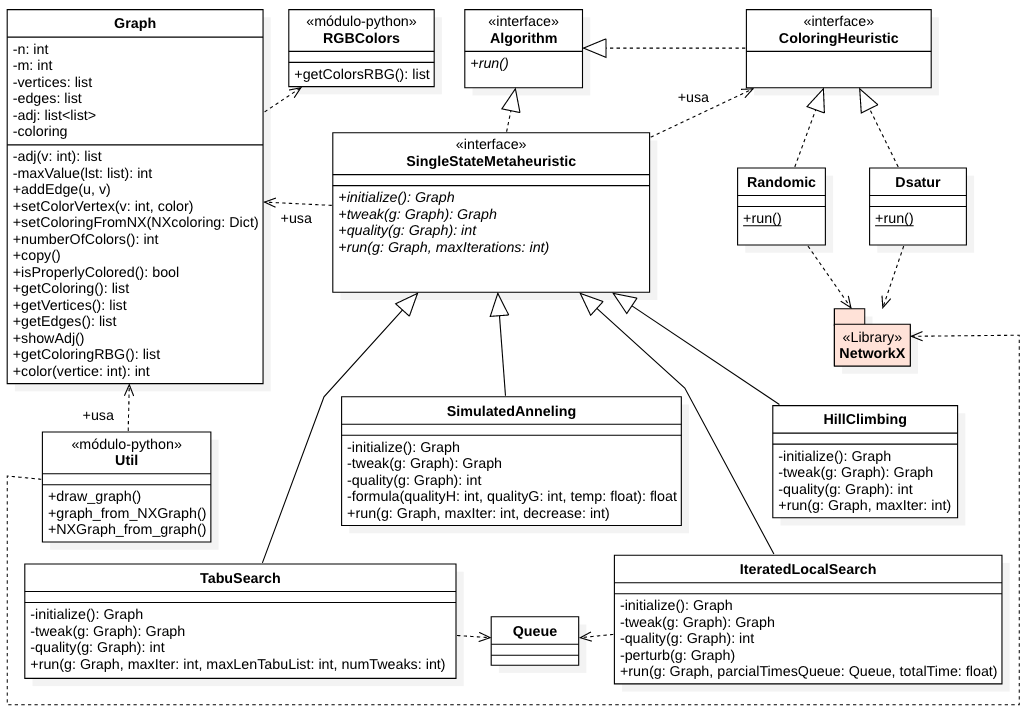}
\caption{Diagrama de classes das meta-heurísticas de estado único.}
\label{fig1}
\end{figure}

As meta-heurísticas de estado único HC, SA, TS e ILS foram modeladas conforme o diagrama de classes da Figura \ref{fig1} e codificadas na linguagem de programação Python 3. Como a implementação trata, essencialmente, de algoritmos, a interface \verb|Algorithm| define o método polimórfico \verb|run()| que é implementado em todas as subclasses concretas de cada meta-heurística. Essa interface é realizada diretamente pela classe abstrata \verb|SingleStateMetaheuristic| que reúne as operações comuns às meta-heurísticas de estado único e que utilizam o grafo (classe \verb|Graph|) sobre o qual as colorações serão aplicadas. A classe \verb|SingleStateMetaheuristic| utiliza as heurísticas de coloração aleatória (classe \verb|Randomic|) e Dsatur (classe \verb|Dsatur|) da bilbioteca python \verb|NetworkX|, por meio da interface \verb|ColoringHeuristic|. Brélaz \cite{Brelaz:1979:NMC:359094.359101} propôs o algoritmo Dsatur que se baseia no conceito de grau de saturação de um vértice (número de cores distintas adjacentes a este vértice) para gerar uma coloração de qualidade superior à coloração aleatória. Na implementação, o Dsatur é utilizado para gerar uma coloração inicial para as meta-heurísticas.

A classe \verb|Graph| implementa o grafo e armazena sua coloração no atributo \verb|coloring| sob a forma de uma lista de cores $[c_1, c_2, \ldots, c_n]$. Cada posição $i$ ($1 \leq i \leq n$) dessa lista corresponde a um vértice do grafo e cada valor uma cor. 
O grafo é implementado sob a forma de uma lista de adjacências no atributo \verb|adj| e contém as operações básicas de inserção de arestas, atribuição de cores à vértices, verificação se a coloração do grafo é própria, e métodos auxiliares. 
A classe \verb|Graph| utiliza o módulo Python \verb|RGBColors| com a função \verb|get_colors_RBG()| que retorna uma lista de 96 cores codificadas no padrão RGB. Na classe \verb|Util| o método \verb|draw_graph()| utiliza a biblioteca \verb|NetworkX| para desenhar o grafo com cores obtidas de \verb|RGBColor|. É importante observar que \verb|draw_graph()| é empregado somente para validar visual a manualmente as soluções das meta-heurísticas para instâncias pequenas do problema com até 10 vértices. O módulo \verb|Util| contém, ainda, métodos de conversão entre a coloração no formato da biblioteca \verb|NetworkX| e a coloração no formato da classe \verb|Graph|, e um método para ler um grafo a partir de um arquivo.

As subclasses \verb|HillClimbing|, \verb|SimulatedAnneling|, \verb|IteratedLocalSearch| e \verb|TabuSearch| implementam a interface \verb|SingleStateMetaheuristic|, cada uma de acordo com a meta-heurística correspondente. A classe auxiliar \verb|Queue| é utilizada por \verb|IteratedLocalSearch| e \verb|TabuSearch|. O método polimórfico \verb|run()| é o único método público destas subclasses e implementa o procedimento principal da meta-heurística.

O código completo das meta-heurísticas pode ser encontrado em \verb|https://github.com/kaninchen-dev|.

%

\subsection{Experimentos}

Os experimentos foram executados em uma máquina virtual com sistema operacional Ubuntu 16.04 LTS (GNU\/Linux 4.4.0-170-generic x86\_64), 20Gb de disco, 16Gb de RAM e quatro processadores, criada a partir de um servidor Dell PowerEdge R820, com 128Gb RAM, 900Gb de HD e Processador Intel\textsuperscript{\textregistered} Xeon\textsuperscript{\textregistered} CPU E5-4617 com 2.90GHz. Utilizou-se o interpretador Python, versão 3.5.2, para a execução dos programas. A Tabela \ref{tab1} apresenta as instâncias de grafos utilizadas nos experimentos obtidas do II Desafio DIMACS \cite{Johnson:1996}.

Inicialmente, cada meta-heurística foi executada com um conjunto inicial de parâmetros para cada uma das instâncias. Estes parâmetros foram empiricamente ajustados até que se obtivesse um tempo razoável de processamento, não ultrapassando o limite de 40h. Após a etapa de ajustes de parâmetros, efetuou-se os experimentos definitivos obtendo-se como resultados os números de cores apresentados na Tabela \ref{tab2} e as respectivas diferenças percentuais entre cores obtidas e cores DIMACS. A Tabela \ref{tab3} apresenta os tempos de execução dos experimentos e a Tabela \ref{tab4} mostra os parâmetros empregados em cada meta-heurística.


\begin{table}[]
\centering
\begin{tabular}{lcccccc}
\hline
~             & Vértices & Arestas & Cores DIMACS \\
\hline
DSJC125.1 & 125 &   736 &  5  \\
DSJC125.5 & 125 &  3891 & 17 \\
DSJC125.9 & 125 &  6961 & 44 \\
\hline
DSJC250.1 & 250 &  3218 &  8 \\
DSJC250.5 & 250 & 15668 & 28 \\
DSJC250.9 & 250 & 27897 & 72 \\
\hline
\end{tabular}
\caption{Instâncias DIMACS utilizadas nos experimentos.}
\label{tab1}
\end{table}

\begin{table}[]
\centering
\begin{tabular}{lcccccccc}
\hline
~             & HC & Dif.\% & SA & Dif.\%  & TS & Dif.\%  & ILS & Dif.\% \\
\hline
DSJC125.1 & 6  & 20,00 & 6  & 20,00 &  6 & 20,00 &  6 & 20,00 \\
DSJC125.5 & 20 & 17,65 & 21 & 23,53 & 21 & 23,53 & 21 & 23,53 \\
DSJC125.9 & 46 &  4,55 & 47 &  6,82 & 47 &  6,82 & 46 &  4,55 \\
\hline
DSJC250.1 & 10 & 25,00 & 10 & 25,00 & 10 & 25,00 & 10 & 25,00 \\
DSJC250.5 & 35 & 25,00 & 36 & 28,57 & 36 & 28,57 & 36 & 28,57 \\
DSJC250.9 & 81 & 12,50 & 83 & 15,28 & 83 & 15,28 & 83 & 15,28 \\
\hline
\end{tabular}
\caption{Números de cores obtidos e suas diferenças percentuais relativas às cores DIMACS.}
\label{tab2}
\end{table}

\begin{table}[]
\centering
\begin{tabular}{lcccccc}
\hline
~             & HC & SA & TS & ILS \\
\hline
DSJC125.1 & 4,55min  &   10,98s  &   2,95s   & 1,07h \\
DSJC125.5 &   1,19h  &  3,34min  &  54,73s   & 1,25h \\
DSJC125.9 &   3,71h  & 10,65min  & 2,92min   & 1,22h \\
\hline
DSJC250.1 & 50,17min &   2,39min &   53,52s  & 1,03h \\
DSJC250.5 & 11,53h   &  39,65min & 17,49min  & 1,46h \\
DSJC250.9 & 31,36h   &     2,44h & 59,49min  & 1,94h \\
\hline
\end{tabular}
\caption{Tempos de execução de cada meta-heurísticas.}
\label{tab3}
\end{table}

\begin{table}[]
\centering
\begin{tabular}{ll}
\hline
~ & Configurações \\
\hline
HC  & iterações = 5.000  \\ 
~ & ~ \\ 
SA  & iterações = 10.000 \\ 
~   & decréscimo = 0.005 \\ 
~ & ~ \\ 
TS  & iterações = 10 \\ 
~   & compr. lista tabu = 20 \\ 
~   & núm. tweaks = 10 \\ 
~ & ~ \\ 
ILS & start = 10s, stop = 100s \\ 
~   & compr. fila = 70 \\ 
\hline
\end{tabular}
\caption{Configurações de parâmetros das meta-heurísticas.}
\label{tab4}
\end{table}

A seção seguinte discute os resultados dos experimentos.

\section{Resultados}

Os experimentos mostram que nenhum dos métodos conseguiu superar os resultados da literatura, dentro dos parâmetros utilizados e do limite de tempo estabelecido. Considerando o fato das instâncias terem ordem e densidades crescentes, esperava-se tempos crescentes de processamento para todas as meta-heurísticas. No entanto, a instância DSJC250.1 processou em menor tempo do que DSJC125.9, e \textit{Iterated Local Search} consumiu tempo aproximadamente constante para todas as instâncias dos experimentos. Ainda assim, o grupo de instâncias DSJC125 exigiu, em média, menor tempo de processamento do que o grupo de instâncias DSJC250. \textit{Hill Climbing} obteve os melhores resultados em número de cores, mas perdeu em tempo de execução, chegando a 31,35h de processamento. Na prática, as demais meta-heurísticas obtiveram os mesmos valores de número de cores (com exceção de \textit{Iterated Local Search} que obteve uma cor a menos para \verb|DSJC125.9|). Interessante notar que as menores diferenças percentuais ocorrem com as instâncias mais difíceis dos grupos \verb|DSJC125| e \verb|DSJC250|. \textit{Hill Climbing} e \textit{Iterated Local Search} obtiveram os números de cores mais próximos dos valores DIMACS para \verb|DSJC125.9|, com diferença percentual de 4,55\% em relação ao benchmark e \textit{Hill Climbing} obteve o melhor resultado para a instância \verb|DSJC250.9|.
Observando a Tabela \ref{tab3}, \textit{Hill Climbing}, \textit{Simulated Annealing} e \textit{Tabu Search} tem um tempo crescente de processamento à medida em que as instância se tornam maiores e mais densas, enquanto o tempo de \textit{Iterated Local Search} se mantém aproximadamente constante.

\section{Discussão}

Este trabalho apresentou um estudo comparativo da eficiência de tempo das meta-heurísticas \textit{Hill Climbing}, \textit{Simulated Annealing} e \textit{Tabu Search} e \textit{Iterated Local Search}. Observa-se que as meta-heurísticas cumprem o seu papel em obter soluções aproximadas cujos valores podem ser melhorados mediante: (1) ajustes em seus parâmetros; (2) maior disponibilidade de tempo para processamento; (3) maior poder computacional da plataforma de execução; (4) implementação de paralelismo. Os ajustes de parâmetros forçam uma exploração mais ampla do espaço de busca, mas podem demandar uma quantidade consideravelmente maior de tempo o que pode tornar inviável o uso da técnica, como demonstram as 31,35h de processamento de \textit{Hill Climbing} (durante os testes, \textit{Tabu Search} processou por três dias utilizando 100 iterações).

Os experimentos demonstram existir um grau de equivalência entre as técnicas relativa ao seu potencial para obter boas soluções. Certamente, \textit{Hill Climbing} destacou-se por ter tido maior tempo de processamento, embora as demais técnicas possam ser configuradas para, igualmente, alcançar melhores resultados. \textit{Iterated Local Search} destacou-se por seu desempenho constante, não obstante a variação das instâncias do problema. Este estudo verificou que estas quatro meta-heurísticas são boas alternativas quando não é possível ou necessário a obtenção de soluções ótimas para problemas intratáveis.

\section{Agradecimentos}

Agradeço à Universidade do Estado do Amazonas que financiou este Estudo Comparativo por meio do Projeto de Produtividade Acadêmica, processo n\textordmasculine~347418, portaria n\textordmasculine~1112/2014.



\bibliographystyle{acm}
\bibliography{sbc-template}

\end{document}